\input harvmac.tex
\noblackbox
%


\def\unlockat{\catcode`\@=11}
\def\lockat{\catcode`\@=12}

\unlockat

\def\newsec#1{\global\advance\secno by1\message{(\the\secno. #1)}
\global\subsecno=0\global\subsubsecno=0\eqnres@t\noindent
{\bf\the\secno. #1}
\writetoca{{\secsym} {#1}}\par\nobreak\medskip\nobreak}
\global\newcount\subsecno \global\subsecno=0
\def\subsec#1{\global\advance\subsecno
by1\message{(\secsym\the\subsecno. #1)}
\ifnum\lastpenalty>9000\else\bigbreak\fi\global\subsubsecno=0
\noindent{\it\secsym\the\subsecno. #1}
\writetoca{\string\quad {\secsym\the\subsecno.} {#1}}
\par\nobreak\medskip\nobreak}
\global\newcount\subsubsecno \global\subsubsecno=0
\def\subsubsec#1{\global\advance\subsubsecno by1
\message{(\secsym\the\subsecno.\the\subsubsecno. #1)}
\ifnum\lastpenalty>9000\else\bigbreak\fi
\noindent\quad{\secsym\the\subsecno.\the\subsubsecno.}{#1}
\writetoca{\string\qquad{\secsym\the\subsecno.\the\subsubsecno.}{#1}}
\par\nobreak\medskip\nobreak}

\def\subsubseclab#1{\DefWarn#1\xdef
#1{\noexpand\hyperref{}{subsubsection}%
{\secsym\the\subsecno.\the\subsubsecno}%
{\secsym\the\subsecno.\the\subsubsecno}}%
\writedef{#1\leftbracket#1}\wrlabeL{#1=#1}}
\lockat

%
%
%

\def\CK{{\cal K}}

\def\CN{{\cal N}}

\def\IZ{\relax\ifmmode\mathchoice
{\hbox{\cmss Z\kern-.4em Z}}{\hbox{\cmss Z\kern-.4em Z}}
{\lower.9pt\hbox{\cmsss Z\kern-.4em Z}}
{\lower1.2pt\hbox{\cmsss Z\kern-.4em Z}}\else{\cmss Z\kern-.4em
Z}\fi}
\def\IB{\relax{\rm I\kern-.18em B}}
\def\IC{{\relax\hbox{$\inbar\kern-.3em{\rm C}$}}}
\def\ID{\relax{\rm I\kern-.18em D}}
\def\IE{\relax{\rm I\kern-.18em E}}
\def\IF{\relax{\rm I\kern-.18em F}}
\def\IG{\relax\hbox{$\inbar\kern-.3em{\rm G}$}}
\def\IGa{\relax\hbox{${\rm I}\kern-.18em\Gamma$}}
\def\IH{\relax{\rm I\kern-.18em H}}
\def\II{\relax{\rm I\kern-.18em I}}
\def\IK{\relax{\rm I\kern-.18em K}}
\def\IP{\relax{\rm I\kern-.18em P}}

\def\inbar{\,\vrule height1.5ex width.4pt depth0pt}
\def\p{\partial}

\font\cmss=cmss10 \font\cmsss=cmss10 at 7pt
\def\IR{\relax{\rm I\kern-.18em R}}

\def\Tr{\rm Tr}

%
%

\lref\fhsv{
S. Ferrara, J. A. Harvey, A. Strominger, C. Vafa ,
``Second-Quantized Mirror Symmetry, '' Phys. Lett. {\bf B361} (1995) 59;
hep-th/9505162. }

\lref\vaftest{C. Vafa, ``A stringy test of the fate of the conifold,''  Nucl.
Phys.
{\bf B447} (1995) 252; hep-th/9505053.}

\lref\hm{J. Harvey and G. Moore,
``Algebras, BPS states, and strings,''
hep-th/9510182; Nucl. Phys. {\bf B463}(1996)315.}

\lref\kv{S. Kachru and C. Vafa, ``Exact results for $N=2$ compactifications of
heterotic strings, '' Nucl. Phys. {\bf B450} (1995) 69, hep-th/9505105. }

\lref\swa{N. Seiberg and E. Witten, ``Electric-magnetic duality, monopole
condensation, and
confinement in $N=2$ supersymmetric Yang-Mills theory,'' Nucl. Phys. {\bf B426}
(1994) 19; (E) {\bf B340} (1994) 485, hep-th/9407087. }

\lref\swb{N. Seiberg and E. Witten, `` Monopoles, duality and chiral symmetry
breaking in $N=2$ supersymmetric QCD, '' Nucl. Phys. {\bf B431} (1994) 484,
hep-th/9408099. }

\lref\dabh{A. Dabholkar and J. A. Harvey, ``Nonrenormalization of the
superstring
tension,'' Phys. Rev. Lett. {\bf 63} (1989) 478; A. Dabholkar, G. Gibbons,
J. A. Harvey and F. Ruiz Ruiz, ``Superstrings and solitons,''
Nucl. Phys. {\bf B340} (1990) 33.}

\lref\sschwarz{A. Sen and J. Schwarz, ``Duality symmetries of 4-D heterotic
strings,''
Phys. Lett. {\bf B312} (1993) 105, hep-th/9305185.}

\lref\hullt{C. Hull and P. Townsend, ``Unity of superstring dualities,'' Nucl.
Phys.
{\bf B438} (1995) 109; hep-th/9410167.}

\lref\wittdyn{E. Witten, ``String theory dynamics in various dimensions,''
Nucl.
Phys. {\bf B443} (1995) 85, hep-th/9503124.}

\lref\klti{A. Klemm, W. Lerche and S. Theisen, ``Nonperturbative effective
actions of
$N=2$ supersymmetric gauge theories,'' hepth-9505150. }

\lref\nsi{S. Cecotti, P. Fendley, K. Intriligator and C. Vafa, ``A new
supersymmetric
index, '' Nucl. Phys. {\bf B386} (1992) 405, hep-th/9204102;
S. Cecotti and C. Vafa, ``Ising model
and $N=2$ supersymmetric theories, '' Commun.  Math. Phys. {\bf 157} (1993)
139,
hep-th/9209085.}

\lref\walton{M.  A. Walton,  ``Heterotic string on the simplest Calabi-Yau
manifold and
its orbifold limit, '' Phys. Rev. {\bf D37} (1988) 377. }

\lref\dkl{L. Dixon,  V. S. Kaplunovsky and J. Louis, ``Moduli-dependence of
string
loop corrections to gauge coupling constants, ''Nucl. Phys. {\bf B329} (1990)
27. }

\lref\vadim{V. Kaplunovsky, ``One loop threshold effects in
string unification,'' Nucl. Phys. {\bf B307} (1988) 145,  revised
in hep-th/9205070.}

\lref\kaplouis{V. Kaplunovsky and J. Louis, ``On gauge couplings in string
theory,''
Nucl. Phys. {\bf B444} (1995) 191, hep-th/9502077. }

\lref\kaplouisp{V. Kaplunovsky and J. Louis, ``Field dependent
gauge couplings in locally supersymmetric
effective quantum field theories,'' Nucl. Phys. {\bf B422} (1994) 57;
hep-th/9402005.}

\lref\cv{E. Calabi and E. Vesentini, Ann. Math. {\bf 71} (1960) 472.}

\lref\FMS{D. Friedan, E. Martinec,  and S. Shenker,
``Conformal Invariance, Supersymmetry, and String Theory,''
Nucl.Phys. {\bf B271} (1986) 93.}

\lref\gilmore{R. Gilmore, {\it Lie Groups, Lie Algebras and some of their
Applications, }Wiley-Interscience, New York, 1974.}

\lref\patera{ J. Patera, R. T. Sharp and P. Winternitz, J. Math. Phys. {\bf 17}
(1976) 1972. }

\lref\agntop{I. Antoniadis,
 E. Gava, K.S. Narain and T.R. Taylor,
``Topological amplitudes in string theory,''
Nucl. Phys. {\bf B413}(1994) 162.}

\lref\agn{I. Antoniadis,  E. Gava, K.S. Narain, ``Moduli corrections to
gravitational
couplings from string loops,'' Phys. Lett. {\bf B283} (1992) 209,
hep-th/9203071; `` Moduli corrections to gauge and gravitational couplings in
four-dimensional superstrings,'' Nucl. Phys. {\bf B383} (1992) 109,
hep-th/9204030.}

\lref\agnt{I. Antoniadis,
 E. Gava, K.S. Narain and T.R. Taylor,
``Superstring threshhold corrections to
Yukawa couplings,'' Nucl. Phys {\bf B407} (1993) 706;
hep-th/9212045. Note: These versions are
different. }

\lref\kennati{See K. Intriligator and N. Seiberg, ``Lectures on supersymmetric
gauge
theories and electric-magnetic duality,'' hep-th/9509066 for a recent review.}

\lref\fvanp{S. Ferrara and A. Van Proeyen, `` A theorem on $N=2$ special
Kahler product manifolds,'' Class. Quantum Grav. {\bf 6}
(1989) L243. }

\lref\recent{Recent test of duality refs}

\lref\effcomp{  C .  Vafa, ``Evidence for F-Theory,'' hep-th/9602022;
 D. R. Morrison,  C. Vafa, ``Compactifications of F-Theory on
Calabi--Yau Threefolds -- I,'' hep-th/9602114;
D. R. Morrison,  C. Vafa, ``Compactifications of F-Theory on
Calabi--Yau Threefolds -- II,'' hep-th/9603161.}

\lref\yoshii{H. Yoshii, ``On moduli space of $c=0$ topological conformal
field theories,'' Phys. Lett. {\bf B275} (1992) 70.}

\lref\nojiri{S. Nojiri, ``$N=2$ superconformal topological field theory,''
 Phys. Lett. {\bf 264B}(1991)57. }

\lref\berkvaf{N. Berkovits and C. Vafa, ``$N=4$
topological strings, ''  Nucl. Phys. {\bf B433} (1995) 123, hep-th/9407190.}

\lref\vwdual{ C. Vafa and E. Witten, ``Dual string pairs with $N=1$ and $N=2$
supersymmetry in four dimensions, '' hep-th/9507050. }

\lref\monstref{B. H. Lian and S. T. Yau, ``Arithmetic properties of mirror map
and quantum
coupling, '' hep-th/9411234. }

\lref\lianyau{B. H. Lian and S. T. Yau, ``Arithmetic properties of mirror map
and quantum
coupling, '' hep-th/9411234;
``Mirror Maps, Modular Relations
and Hypergeometric Series I ,'' hep-th/9506210;
``Mirror Maps, Modular
Relations and Hypergeometric Series II,''
hep-th/9507153}

\lref\bcov{M. Bershadsky, S. Cecotti, H. Ooguri and C. Vafa, `` Kodaira-Spencer
theory
of gravity and exact results for quantum string amplitudes, '' Commun. Math.
Phys.
{\bf 165} (1994) 311, hep-th/9309140. }

\lref\fgz{I. Frenkel, H. Garland, and G. Zuckerman, ``Semi-infinite
cohomology and string theory,'' Proc. Nat. Acad. Sci.
{\bf 83}(1986) 8442.}

\lref\gross{ D. Gross, ``High energy symmetries of
string theory,'' Phys. Rev. Lett. {\bf 60B} (1988) 1229.}

\lref\horava{P. Horava, ``Strings on world sheet orbifolds,''  Nucl. Phys.
{\bf B327} (1989) 461; ``Background duality of open string models,'' Phys.
Lett.
{\bf B231}(1989)251.}

\lref\sgntti{M. Bianchi, G. Pradisi and A. Sagnotti, ``Toroidal
compactification and symmetry breaking in open string theories,''
 Nucl. Phys. {\bf B376} (1992) 365.}

\lref\polch{J. Polchinski, ``Combinatorics of boundaries in
string theory,'' Phys. Rev. {\bf D50} (1994) 6041,
hep-th/9407031;  M. B. Green,
``A gas of D instantons,'' Phys. Lett. {\bf B354} (1995) 271,
hep-th/9504108.}

\lref\shenker{S. Shenker, ``Another Length Scale in
String Theory,'' hep-th/9509132}

\lref\argf{P. C. Argyres and A. E.  Faraggi, ``The vacuum structure and
spectrum of
$N=2$ supersymmetric $SU(n)$ gauge theory, '' Phys. Rev. Lett.
{\bf 74} (1995) 3931, hep-th/9411057.}

\lref\givpor{A. Giveon and M. Porrati, ``Duality invariant string algebra and
$D=4$ effective actions, '' Nucl. Phys. {\bf B355} (1991) 422. }

\lref\dfkz{J. P. Derendinger, S. Ferrara, C. Kounnas and F. Zwirner, ``On loop
corrections to string effective field theories: field-dependent gauge couplings
and
$\sigma$-model anomalies,'' Nucl. Phys. {\bf B372} (1992) 145. }

\lref\agnti{I. Antoniadis, E. Gava, K. S. Narain and T. R. Taylor, ``$N=2$ Type
II-
Heterotic duality and higher derivative F-terms, '' hep-th/9507115.}

\lref\klt{V. Kaplunovsky, J. Louis and S. Theisen, ``Aspects of duality in
$N=2$ string vacua,'' Phys. Lett. {\bf B357} (1995) 71, hep-th/9506110.}

\lref\louispas{J. Louis, PASCOS proceedings, P. Nath ed., World
Scientific 1991.}

\lref\lco{G. L. Cardoso and B. A. Ovrut, ``A Green-Schwarz mechanism
for $D=4$, $N=1$ supergravity anomalies,'' Nucl. Phys. {\bf B369} (1992) 351;
``Coordinate and Kahler sigma model anomalies and their cancellation
in string effective field theories,''  Nucl. Phys. {\bf B392} (1993) 315,
hep-th/9205009.}

\lref\levin{L. Levin, Polylogarithms and Associated Functions,
North Holland 1981. See eq.  6.7.}

\lref\witthyper{E. Witten, ``Topological tools in ten dimensional physics, ''
in
{\it Unified String Theories}, eds. M. Green and D. Gross, World Scientific,
Singapore,
1986.}

\lref\rey{ G. L. Cardoso,  G. Curio,  D. Lust,  T.
Mohaupt,  S.-J. Rey, ``BPS Spectra and Non--Perturbative Couplings in
N=2,4 Supersymmetric String Theories,'' Nucl. Phys. {\bf B464} (1996) 18;
hep-th/9512129.}

\lref\fein{A. Feingold and I. Frenkel,
``A Hyperbolic Kac-Moody Algebra and the
Theory of Siegel Modular Forms of Genus
2,'' Math. Ann. {\bf 263} (1083) 87.}

\lref\golatt{P. Goddard and D. Olive, ``Algebras, Lattices,
and Strings,''
in {\it Vertex operators in mathematics and
physics},'' ed. J. Lepowsky et. al. Springer-Verlag, 1985.}

\lref\gebnic{R. W. Gebert and H. Nicolai, `` On $E_{10}$ and the DDF
construction,''
hep-th/9406175.}

\lref\wittorb{E. Witten, ``Space-time and topological orbifolds,'' Phys. Rev.
Lett.
{\bf 61} (1988) 670.}

\lref\fuchs{J. Fuchs, {\it Affine lie algebras and quantum groups,} Cambridge
University Press,  Cambridge, 1992. }

\lref\cfetc{S. Cecotti, S. Ferrara, L. Girardello, A. Pasquinucci , M. Porrati
, ``Matter coupled supergravity with
Gauss-Bonnet invariants: Component Lagrangian
and supersymmetry breaking,'' Int. J. Mod. Phys. {\bf A3} (1988)
1675}

\lref\kac{V. G. Kac, {\it Infinite dimensional Lie algebras,} Cambridge
University Press, Cambridge, 1990. }

\lref\moonshine{J. H. Conway and S. P. Norton, ``Monstrous moonshine,''
Bull. London Math. Soc. {\bf 11} (1979) 308. }

\lref\nikulin{see e.g. V. V. Nikulin, ``Reflection groups in hyperbolic spaces
and the denominator formula for Lorentzian Kac-Moody algebras,
alg-geom/9503003.}

\lref\kmw{V. G. Kac, R. V. Moody, and M. Wakimoto, ``On $E_{10}$,'' In K.
Bleuler
and M. wener, eds. {\it Differential geometrical methods in theoretical
physics.}
Proceedings, NATO advanced research workshop, 16th international conference,
Como
Kluwer, 1988.}

\lref\gebnici{R. W. Gebert and H. Nicolai, ``$E_{10}$ for beginners,''
hep-th/9411188.}

\lref\flm{I. B. Frenkel, J. Lepowsky, and A. Meurman, {\it Vertex operator
algebras
and the monster,} Pure and Applied Mathematics Volume 134, Academic
Press, San Diego, 1988.}

\lref\lianzuck{B. H. Lian and  G. J. Zuckerman, ``New
Perspectives on the BRST Algebraic Structure of
String Theory,''  Commun.Math.Phys. {\bf 154} (1993) 613,
hep-th/9211072.}

\lref\mool{C. Montonen and D. Olive, ``Magnetic monopoles as
gauge particles? ''Phys. Lett. {\bf 72B} (1977)
117; P. Goddard, J. Nuyts and D. Olive, ``Gauge theories
and magnetic charge,''  Nucl. Phys. {\bf B125} (1977) 1.}

\lref\WO{E. Witten and D. Olive, ``Supersymmetry algebras that
include topological charges,'' Phys. Lett. {\bf 78B} (1978) 97. }

\lref\senb{A. Sen, ``Dyon-monopole bound states, selfdual harmonic
forms on the multi-monopole moduli space, and $SL(2,Z)$ invariance
in string theory,'' Phys. Lett. {\bf 329} (1994) 217,
hep-th/9402032.}

\lref\osborn{H. Osborn, ``Topological charges for
$N=4$ supersymmetric gauge theories and monopoles of
spin 1,'' Phys. Lett. {\bf 83B} (1979) 321.}

\lref\andycone{A. Strominger, ``Massless black holes and conifolds in
string theory,'' Nucl. Phys. {\bf B451} (1995) 96; hep-th/9504090.}

\lref\coneheads{B. R. Greene, D. R. Morrison and A. Strominger, ``Black hole
condensation and the unification of string vacua,'' Nucl. Phys. {\bf B451}
(1995) 109; hep-th/9504145}

\lref\joebrane{J. Polchinski, ``Dirichlet-Branes and Ramond-Ramond charges''
hep-th/9510017.}

\lref\sena{A. Sen, ``Strong-weak coupling duality in four-dimensional string
theory,'' Int. J. Mod. Phys. {\bf A9} (1994) 3707.}

\lref\givrev{A. Giveon, M. Porrati and E. Rabinovici, ``Target space duality in
string
theory,'' Phys. Rep. {\bf 244} (1994) 77;hep-th/9401139}

\lref\kirka{E. Kiritsis and C. Kounnas, ``Infrared Regularization of
superstring
theory and the one-loop calculation of coupling constants,'' Nucl. Phys.
{\bf B442} (1995) 442, hep-th/9501020.}

\lref\min{J. Minahan, Nucl. Phys. {\bf B298} (1988) 36.}

\lref\afgntrev{I. Antoniadis, S. Ferrara, E. Gava, K. S. Narain and T. R.
Taylor,
``Duality symmetries in $N=2$ heterotic superstring,'' hep-th/9510079.}

\lref\bjulia{B. Julia, in {\it Applications of group theory in physics
and mathematical physics,} ed. P. Sally et. al. (American Mathematical
Society, Providence, 1985). }

\lref\dwvp{B. de Wit and A. Van Proeyen, ``Broken
sigma-model isometries in very special
geometry,'' Phys. Lett. {\bf 293B}(1992)94.}

\lref\klm{A. Klemm, W. Lerche and P. Mayr, ``K3-fibrations and
Heterotic-Type II string duality, '' Phys. Lett. {\bf B357}
(1995) 313, hep-th/9506122.}

\lref\cogp{P. Candelas, X. de la Ossa, P. S. Green and L. Parkes,
``A pair of Calabi-Yau manifolds as an exactly soluble
superconformal theory,'' Nucl. Phys. {\bf B359} (1991) 21.}

\lref\hm{J. A. Harvey and G. Moore, ``Algebras, BPS states, and strings,''
Nucl. Phys. {\bf B463} (1996) 315; hep-th/9510182.}

\lref\nforefs{refs on $n=4$ SUGRA, berghshoeff, de Roo etc.}

\lref\chs{C. G. Callan, J. A. Harvey and A. Strominger, Nucl. Phys.
{\bf B359} (1991) 611; C. G. Callan, J. A. Harvey and A. Strominger,
Nucl. Phys. {\bf 367} (1991) 60.}

\lref\seiberg{N. Seiberg, Phys. Lett. {\bf B206} (1988) 75.}

\lref\dufflu{M. J. Duff and J. X. Lu, ``Elementary five-brane solutions
of $D=10$ supergravity,'' Nucl. Phys. {\bf B354} (1991) 141.}

\lref\andybrane{A. Strominger, ``Heterotic Solitons,'' Nucl. Phys.
{\bf B343} (1990) 167; ERRATUM ibid. {\bf B353} (1991) 565.}

\lref\wittensmall{E. Witten, ``Small instantons in string theory,'' Nucl.
Phys. {\bf B460} (1996) 541; hep-th/9511030.}

\lref\horwitta{P. Horava and E. Witten, Nucl. Phys. {\bf B460} (1996)
506; hep-th/9510209.}

\lref\horwittb{P. Horava and E. Witten, ``Eleven-dimensional supergravity
on a manifold with boundary,'' Nucl. Phys. {\bf B475} (1996) 94;
hep-th/9603142.}

\lref\wittenexp{E. Witten, ``Strong coupling expansion of Calabi-Yau
compactification,'' hep-th/9602070.}

\lref\edtob{E. Witten, ``Some comments on string dynamics,''
hep-th/9507121.}

\lref\andyopen{A. Strominger, ``Open p-branes,'' Phys. Lett. {\bf B383}
(1996) 44; hep-th/9512059.}

\lref\gh{O. J. Ganor and A. Hanany, ``Small E(8) instantons and tensionless
noncritical strings,'' Nucl. Phys. {\bf B474} (1996) 122; hep-th/9602120.}

\lref\wip{Work in progress.}

\lref\bcova{M. Bershadsky, S. Cecotti, H. Ooguri and C. Vafa,
``Holomorphic anomalies in topolgical field theories,''Nucl. Phys.
{\bf B405} (1993) 279; hep-th/9302103.}

\lref\dealwis{S. P. de Alwis, ``A note on brane tension and $M$-theory,''
hep-th/9607011.}

\lref\carlustov{G.L. Cardoso, D. Lust, and
B.A. Ovrut, ``Moduli dependent non-holomorphic
contributions of massive states to gravitational
couplings and $C^2$ terms in $Z(N)$ orbifold compactification,''
Nucl. Phys. {\bf B436} (1995) 65; hep-th/9410056.}

\lref\hs{J. A. Harvey and A. Strominger, Nucl. Phys. {\bf B458} (1996) 456;
hep-th/9504047.}

\lref\bderoo{E. A. Bergshoeff and M. de Roo, Nucl. Phys. {\bf B328} (1989)
439.}

\lref\instrefs{D. Finnell and P. Pouliot, Nucl. Phys. {\bf B453} (95) 225,
hep-th/9503115; K. Ito and N. Sasakura, Phys. Lee. {\bf B382} (1996) 95,
hep-th/9602073;  N. Dorey, V. W. Khoze and M. P. Mattis, Phys. Rev. {\bf D54}
(1996) 2921, hep-th/9603136.}

\lref\vw{C. Vafa and E. Witten, ``A strong coupling test of S-duality,''
Nucl. Phys. {\bf B431} (1994) 3; hep-th/9408074.}

\lref\swts{N. Seiberg and E. Witten, ``Comments on string dynamics in
six-dimensions,'' Nucl. Phys. {\bf B471} (1996) 121.}

\lref\bdw{E. Bershoeff, de Roo,
B. de Wit, Nucl. Phys. {\bf B182} (1981)173}

\lref\vwdual{ C. Vafa and E. Witten, ``Dual string pairs with $N=1$ and $N=2$
supersymmetry in four dimensions, '' hep-th/9507050. }

\lref\ssinv{J. H. Schwarz and A. Sen, ``Duality symmetric actions,''
Nucl. Phys. {\bf B411} (1994) 35; hep-th/9304154.}

\lref\sensol{A. Sen, ``String string duality conjecture in six-dimensions and
charged solitonic strings,'' Nucl. Phys. {\bf B450} (1995) 103;
hep-th/9504027.}

\lref\bepaper{J. Harvey and G. Moore,
``Exact gravitational threshhold correction in the
FHSV model,'' to appear.}

%
%

\Title{\vbox{\baselineskip12pt
\hbox{hep-th/9610237}
\hbox{EFI-96-38}
\hbox{YCTP-P21-96 }
}}
{\vbox{\centerline{Fivebrane Instantons}
\centerline{and}
\centerline{ $R^2$ couplings in  $N=4$ String Theory
 } }}

\centerline{Jeffrey A. Harvey}
\bigskip
\centerline{\sl Enrico Fermi Institute, University of Chicago}
\centerline{\sl 5640 Ellis Avenue, Chicago, IL 60637 }
\centerline{\it harvey@poincare.uchicago.edu}
\bigskip
\centerline{Gregory Moore}
\bigskip
\centerline{\sl Department of Physics, Yale University}
\centerline{\sl New Haven, CT  06511}
\centerline{ \it moore@castalia.physics.yale.edu }

\bigskip
\centerline{\bf Abstract}

We compute the gravitational coupling $F_1$ for IIA string
theory on $K3 \times T^2$ and use string-string duality to deduce
the corresponding term for heterotic string on $T^6$. The
latter is an infinite sum of gravitational instanton effects
which we associate with the effects of Euclidean fivebranes
wrapped on $T^6$. These fivebranes are the neutral fivebranes
or zero size instantons of heterotic string theory.

\Date{October 28, 1996}
%

\newsec{Introduction}

The principle of second quantized mirror symmetry \fhsv\ allows
one to map world-sheet instanton effects in compactifications of
$IIA$ string theory to spacetime
instanton effects in dual heterotic string theories.  For the most
part this has been studied in $N=2$ dual pairs \refs{\kv, \fhsv} and the
non-perturbative effects deduced in the heterotic string
in this way can be attributed to Yang-Mills instanton effects, suitably
dressed up by string theory.

In this note we will study this
phenomenon in the much simpler context of the $N=4$ dual pair
consisting of the IIA string on $K3 \times T^2$ and the heterotic string
on $T^6$ \hullt. By studying purely gravitational couplings we will be
able to map genus one world sheet instanton effects on the IIA side
to gravitational instanton effects on the heterotic side. These
instantons are the neutral fivebranes or zero size gauge instantons
of heterotic string theory \refs{\dufflu, \chs, \wittensmall}.
These configurations have many dual descriptions. For example in M theory
this fivebrane can be viewed as the zero
size instanton of M theory which sits at the intersection of the
Coulomb and Higgs branches of the M theory fivebrane moduli space.
We call it a gravitational instanton since the gauge fields vanish
in the corresponding solution of the low-energy field theory
and the fermion zero modes involve
the gravitino and dilatino but not the gaugino fields.

\newsec{Curvature-squared couplings for
the $IIA$ string on $K3 \times T^2$}

Much effort has been devoted to the study of special higher
derivative F terms in string theory with $N=2$ spacetime supersymmetry.
As shown in  \agntop\ these $F$ terms are
related to the topological amplitudes $F_g$
studied in \bcov.

While the $F_g$ have been much studied in
$N=2$ compactifications, in fact, they are
not completely trivial in $N=4$ compactifications.
In this paper we study the first of these quantities, $F_1$,
in the simpler context of the $N=4$ dual pair of IIA string
theory on $K3 \times T^2$ and the heterotic string on $T^6$.
{}From a mathematical point of view the computation is
rather trivial. From a physical point of view, it is not.
In a companion paper we will consider a closely
related $N=2$ dual pair for which the nonperturbative
$F_1$ can be written exactly \bepaper.

Returning to $N=4$ theory, we
 first consider the IIA side. The $T^2$ has moduli $(T,U)$
which are the complexified Kahler modulus and complex structure
modulus of $T^2$ respectively. The global moduli space on $K3 \times T^2$
takes the form
\eqn\twomod{\left( O(22,6;\IZ)\backslash O(22,6;\IR ) / [ O(22) \times O(6) ]
\right)
\times \left( Sl(2,\IZ) \backslash Sl(2; \IR) / U(1) \right) }
where the final factor is associated to the
K\"ahler modulus $T$.
\foot{In this section we use automorphic conventions with ${\rm Im} T >0$.}
The moduli parameterizing the first factor are the
K3 $\sigma$-model moduli, the IIA dilaton,   the
complex structure  modulus $U$ of $T^2$,
and the Wilson lines on $T^2$ of the RR gauge fields.

\subsec{Computation of $F_1$}

The quantity $F_1$ in IIA string theory  is defined
  as a fundamental domain
integral following \bcov:
%
\eqn\bcoviii{
F_1 \equiv  \int_{\CF} {d^2 \tau \over \tau_2}
\Biggl[
\Tr_{R,R} (-1)^{J_L} J_L  (-1)^{J_R} J_R
q^H \bar q^{\tilde H} - const\Biggr]
}
where the trace is over the Ramond-Ramond sector of the internal
superconformal algebra (SCA). The constant term is determined by
the massless spectrum and ensures that the integral is convergent.
As in \hm\ one can analyze the states that contribute to $F_1$
by decomposing them under the left and right-moving superconformal
algebras. Only the RR
BPS states in short (but not medium)
representations of the spacetime
$N=4$  supersymmetry algebra contribute.

The integral \bcoviii\ is easily evaluated for
$K3 \times T^2$ compactifications.
 Both the left
and right SCA decompose as
\eqn\chrlalg{
\tilde \CA^{N=2}_{\tilde c=3} \oplus \tilde
\CA^{N=4}_{\tilde c=6} \quad .
}
Correspondingly, $J= J^{(1)}+J^{(2)}$
where $J^{(2)} = 2 J^3$ from the
$c=6$   $\CN=(4,4)$ superconformal algebra
with $J^3$ the Cartan generator of an
 $SU(2)$ current algebra
and hence
 $\Tr J^{(2)} (-1)^{J^{(2)}} = 0$. Therefore,  only the
term with $\Tr   (-1)^{J^{(2)}} = \chi(K3) = 24 $ contributes
and
we can write
\bcoviii\ as
\eqn\workbcov{
\eqalign{
F_1 &=  \int_{\CF} {d^2 \tau \over \tau_2}
\Biggl[ \Tr^{(1)}_{R,R} J_L^{(1)} J_R^{(1)}
(-1)^{J_L^{(1)} + J_R^{(1)} }
q^H \bar q^{\tilde H} \Tr_{R,R}^{(2)}
(-1)^{J_L^{(2)} + J_R^{(2)} }
q^H \bar q^{\tilde H}  - 24  \Biggr] \cr
& =-  \int_{\CF} {d^2 \tau \over \tau_2}
\biggl( 24 Z_{\Gamma^{2,2}(T,U)} - 24 \biggr) \cr
& =
  24 \biggl( \log   \parallel \eta^2(T) \parallel^2
+ \log   \parallel \eta^2(U) \parallel^2 -
\log[{8 \pi e^{ 1 - \gamma_E} \over  \sqrt{27}} ] \biggr) \cr}}
where $ \parallel \eta^2(T) \parallel^2 \equiv
Im T \vert  \eta^2(T)  \vert^2$ is the
invariant norm-squared, and
in the last line we have used the
result of \dkl.

Note that \workbcov\ is invariant under the $Sl(2,\IZ)$ group acting on $T$.
However, the expression is not $O(22,6)$ invariant since the complex structure
modulus $U$ mixes into
other moduli in the $O(22,6)$  coset. Of course,
the equations of motion of the low energy effective
theory must be $U$-duality invariant.

\subsec{Relation to the effective action}

It was shown in \refs{\bcova,\bcov } that for a
Calabi-Yau $\sigma$-model, quite generally,
$F_1$
splits as a sum
\eqn\splits{
F_1= F_1^{\rm complex} + F_1^{\rm Kahler}
}
that depend only on complex and
Kahler moduli respectively and are
exchanged by mirror symmetry.
Which of these two functions couples
to $\tr R\wedge R$ depends on whether we
discuss the IIA or IIB theory.
In IIA theory only the term
$F_1^{\rm Kahler}$ in \splits\ appears in the low-energy effective
theory and this term is invariant under both $Sl(2,\IZ)$ and
$O(22,6)$. In the IIB theory we would keep the complex
structure term in \splits\ but the moduli space \twomod\ is
also changed by the interchange of $T$ and $U$.

Supersymmetry constrains the local
Wilsonian couplings of $R^2$ to
the Kahler moduli to be holomorphic.
\foot{
Supersymmetric
completions of terms of the form
$\int z \tr R \wedge R$ have been discussed
extensively in \cfetc\carlustov. When comparing
with these expressions it is important to bear in
mind that string amplitudes are only computed
on-shell.}
Of course, $F_1^{\rm Kahler}$ extracted
from \workbcov\ is not holomorphic.
This amplitude is related to an
effective coupling. Nevertheless, we
may extract from it the holomorphic Wilsonian
coupling to $R^2$.
(The relation of the
nonholomorphy of effective couplings and
the holomorphy constraints of Wilsonian
couplings is  subtle and is discussed at
length in
\refs{\dfkz,\lco,\kaplouis,\kaplouisp,\carlustov}.)
The bosonic
terms in the  Wilsonian
action for the $T$-modulus, including
leading couplings to   gravity
fields are (in Minkowski space):
\eqn\effterm{
\eqalign{
I = & { 1\over  2 \kappa_4^2} \int \sqrt{-g}
{ \p_\mu T \p^\mu \bar T \over  (Im T)^2} +
I^{\rm gauge fields} \cr
+ & {1 \over  16 \pi}  Re  \left[
\int  { \log (\eta(T))^{24} \over 2 \pi i}   \tr (R - i R^*)^2
 \right] \cr}
}
Here $\kappa_4^2 = 1/M_{\rm Planck}^2$.
 The curvature tensor is regarded
as a 2-form with values in the Lie algebra of $SO(3,1)$,
$R = \half R^a_{~~ b \mu \nu} dx^\mu dx^\nu$,
the dual on $R$ is taken on the tangent
space indices  and the
trace is over these indices.

We recall
the coupling to the gauge fields
which follows from  the
general constraints of $d=4, N=4$ supergravity
\ref\deroo{M. de Roo, ``Matter coupling in
$N=4$ supergravity,'' Nucl. Phys. {\bf B255}(1955)515}. The
scalar geometry is fixed to be an
$SL(2,\IR) \times O(6,n)$  coset.
 Following \ssinv\
we may write the action in the present
case by introducing
$U(1)$ gauge field strengths
$F^I$,    (considered as 2-forms), $I=1,\dots, 28$, a
quadratic form $\langle v,w \rangle = v^I L_{IJ} w^J $ defining $O(22,6)$
and  $M_{IJ}$, a matrix of scalar moduli for
the $O(22,6)$ coset such that $M^{T} = M,
(ML)^2=1$. We define projection operators
$\Pi_\pm = \half(1 \pm  ML)$ onto the graviphotons
and vectormultiplet field strengths respectively and also define
$\CF_\epsilon^\eta \equiv \Pi_{\epsilon} (F + \eta i * F) $
with $\epsilon = \pm, \eta = \pm $.

Under $SL(2,\IR)$ $\CF_\epsilon^\eta$ transforms
as a modular form of
weight $(0,1)$ when $\epsilon \eta =1$ and of
weight $(1,0)$ when $\epsilon \eta =-1$:
\eqn\trneff{
\eqalign{
\CF_+^+  & \rightarrow (c \bar T + d)   \Pi_+ (F+ i * F)\cr
\CF_-^+  & \rightarrow (c   T + d)   \Pi_- (F+ i * F). \cr}
}
Moreover $\CF_\epsilon^\eta \rightarrow \Omega \CF_\epsilon^\eta$ under
$O(22,6)$ transformations
$\Omega$. The coupling to gauge fields is:
\eqn\gaugflds{
I^{\rm gauge fields} = -{1 \over  16 \pi}
Re \Biggl\{ \int_{\IR^{1,3} }
  \bar T \bigl[ \langle \CF_+^+, \CF_+^+ \rangle +
\langle \CF_-^-, \CF_-^- \rangle \bigr]\Biggr\}
}

Finally, let us discuss the invariances of
the action \effterm.
As emphasized in \sena,  \gaugflds\ is not
manifestly invariant. This is not surprising
since the gauge fields undergo duality
rotations under $SL(2,\IR)$. On the other hand
the Einstein metric is $SL(2,\IR)$-invariant,
and hence the coupling of $T$ to it must
be invariant. This is the key difference between
the gravity coupling in \effterm\ and the gauge
coupling in \gaugflds.
Actually,  \effterm\ is not exactly invariant  because
$\log (\eta(T))^{24}$ suffers a shift under
$SL(2,\IR)$.  As explained in
\refs{\dfkz,\lco,\kaplouis,\kaplouisp,\carlustov}\
this is closely connected with
$\sigma$-model  duality anomalies.
Indeed, the gravitinos, dilatinos and gauginos
 are chiral under $SL(2,\IR)$. Since
all the fields are neutral under the 28 gauge fields
the anomalous variation will have an imaginary
part proportional only to $\tr R \wedge R$.
The anomalous variation of the fermion
determinant cancels the
shift of $\log (\eta(T))^{24}$.
It is worth emphasizing that the nonholomorphic
terms in $F_1$ are nonzero in this example, even
though the ``gravitational $\beta$-function''
of \agn\ is zero.
\foot{The reader should compare with
 the discussion in \rey.}

It would be very interesting to extend the above
discussion to the higher $F_g$ terms.

\newsec{Curvature-squared couplings for
the heterotic string on $T^6$}

Under six dimensional
string-string duality the $T$ modulus of  the $IIA$ theory
on $T^2\times K3$
is exchanged with
the dilaton-axion multiplet or axiodil $\tau_S \equiv 4 \pi i S$
of the heterotic string on $T^6$.
Thus to obtain the $R^2$ couplings in the heterotic theory we
may simply replace $T \rightarrow \tau_S$
everywhere in the previous section.

It is also easy
to argue for this result directly in the
heterotic string by insisting on $S$-duality.
At tree level the   Bianchi identity for $H$,
which follows from implementing
the Green Schwarz mechanism, requires a term in
the Minkowskian action:
\eqn\gsact{
 {1 \over 8 \pi}\int
 Re(\tau_S) \bigl[ \tr R\wedge R - \tr F\wedge F\bigr]
}
Where as usual the gauge trace is in the fundamental of
$SO(32)$ or $1/30$ times the trace in the adjoint for
$E_8 \times E_8$. The coefficient should be exactly as given, since otherwise
instantons would not break the continuous $SL(2,\IR)$
duality group to $SL(2,\IZ)$.
Supersymmetry then requires the coupling of $S$
to $R^2$ to be:
\eqn\susycom{
{1 \over 8 \pi} \int Re(\tau_S) \tr (R\wedge R) -
Im(\tau_S)  \tr (R \wedge R^*) .
}
{}From
$S$-duality itself, we know that the S-dual
completion must take
\eqn\sdcomp{
\tau_S \rightarrow {24 \over  2 \pi i } \log \eta(\tau_S)
}
which leads to the $R^2$ couplings
\eqn\rsqs{  {1 \over 16 \pi} Re  \left[
 \int  {\log \eta^{24}(\tau_S)  \over 2 \pi i}  \tr (R - i R^*)^2
 \right] }
which reproduces  \effterm\ after exchanging $T$ and $\tau_S$.

In terms of effective couplings we may state
the result in terms of the equations of motion
for the dilaton multiplet:
\eqn\onepoint{
\eqalign{
{ 1\over  2 \kappa_4^2} \biggl[
{\nabla ^2 \tau_S \over (Im \tau_S)^2} + i { \nabla^\mu \tau_S \nabla_\mu
\tau_S
\over  (Im \tau_S)^3}\biggr]
& +  * {1\over  16 \pi}
\Biggl[    \langle \CF_+^+, \CF_+^+ \rangle +
\langle \CF_-^-, \CF_-^- \rangle \cr
&
-    \widehat{E_2}(\bar \tau_S)
\Tr (R_{\mu\nu} + i  R_{\mu\nu}^*)^2\Biggr]
 =0  \cr}
}

The first two terms in \onepoint\ transform
under $SL(2,\IR)$ transformations covariantly with
weight two. The last term breaks the invariance
to $SL(2,\IZ)$. The Eisenstein series
$E_2$ transforms with a shift while
$\widehat{E_2}= E_2 - {3 \over  \pi Im \tau} $
transforms covariantly.
The equations \onepoint\   are the $R^2$-corrections to
the S-duality invariant
equations of \sena.

It is worth noting that in the low-energy field theory limit
where we fix the dilaton $S$ to be constant and work on a
general Euclidean four manifold
\rsqs\  contributes
\eqn\ftred{
\exp\biggl[ - (\chi + {3 \over  2} \sigma) \log \eta^{12}
- ( \chi - { 3 \over  2 } \sigma) \log \bar \eta^{12} \biggr]
}
to the Euclidean path integral.
Here $\chi$ is the Euler character and
$\sigma$ is the signature. We presume that this gravitational S-duality
anomaly
is related to the S-duality anomaly in the gauge partition
function studied in \vw. Note in particular that on a four-dimensional
hyperkahler
manifold where the physical and twisted $N=4$ theories should agree the
curvature is automatically anti-self-dual and as a result $\chi = - 3 \sigma/2$
so that the term discussed here contributes $(\bar \eta)^{-24 \chi}$ to
the Euclidean path integral. It would be interesting to make the connection
to the result of \vw\ more precise.

\newsec{Physical interpretation}

As  mentioned earlier, we expect that
worldsheet instantons effects in the IIA
string should be exchanged
with spacetime instanton effects  in
the heterotic string. The formula
for $F_1$  is, as explained in
\refs{\bcov \bcova}, a sum over genus one IIA
worldsheet instantons. In this section we will
identify the spacetime instanton in the heterotic string
which leads to the $R^2$ corrections \rsqs. Since the heterotic
string and the IIA string can each be viewed as wrapped fivebranes
in the dual theory \refs{\sensol,\hs, \wittensmall},
we expect that the instantons can be viewed
as heterotic fivebranes wrapped on $T^6$.

\subsec{Instanton expansion}

In order to make the instanton
expansion manifest we
use the ``string conventions'' with axion-dilaton
chiral superfield $S$ with $Re(S)>0$ and
$$
q_S = e^{-8 \pi^2 S} = e^{2 \pi i \tau_S}.
$$
We
normalize the four dimensional gauge action to be $(1/2 g^2) \int
{\rm tr} F_{\mu \nu}F^{\mu \nu}$ with ${\rm tr}$ the trace in the fundamental
representation of $U(n)$
so that a charge one instanton has action
$ 8 \pi^2 / g^2$. Then
$S={1 \over  g^2} + i {\theta \over  8 \pi^2}$,
and we can expand:
\eqn\hetexp{ \log (\eta(\tau_S))^{24} =  - 8 \pi^2 S
-24 \left[ q_S + {3 q_S^2 \over 2} + {4 q_S^3 \over 3}
+ \cdots  \right] .}
The first term is the tree level coupling, as
discussed above. The higher order terms have the
form of instanton corrections to the $R^2$ couplings where the
instanton action is given by $8 \pi^2 Re(S) = 8 \pi^2 /g^2$.

\subsec{Wrapped 5branes: Macroscopic analysis}

The result \rsqs\ appears to sum up an
infinite set of instanton contributions. To confirm this we would
like to identify the instanton configurations in the heterotic
string which lead to these corrections to $R^2$ couplings.

We now
argue that the relevant instanton is the neutral fivebrane wrapped
on $T^6$. We will proceed in two steps, first analyzing the instanton
using the low-energy analysis of \chs\ and then discussing
the non-perturbative modifications found in \wittensmall.

As in the   one instanton contribution to the $N=2$ prepotential
\seiberg\ the easiest quantity to calculate is not the purely bosonic
term in the action but rather the term with the maximal number of
fermion fields which is related to
the bosonic term by extended supersymmetry.
In $N=4$ supergravity a coupling of the form $ F(S) {\rm tr} R^2$
is paired with 8 fermion
terms involving the dilatino and gravitino. To see this we note
that such 8 fermion terms are present in $N=1$, $d=10$ supergravity
and are paired with the tree level $S {\rm tr} R^2$ coupling by
supersymmetry \bderoo. They must thus be present in the dimensional
reduction to the $N=4$ theory in $d=4$. There are of course additional
terms with fewer fermion fields, these must be generated by
supersymmetric instanton perturbation theory as has been
checked in detail for $N=2$ gauge theory \instrefs.
We are thus looking for an instanton
in heterotic string theory which has action $e^{-8 \pi^2 S}$
and 8 fermion zero modes constructed
out of the gravitino and dilatino but independent of the gauginos.

In \chs\ a number of fivebrane solutions to heterotic string
theory were discussed, the neutral fivebrane, gauge fivebrane
and symmetric fivebrane
\foot{The gauge fivebrane was first discussed
in \andybrane.  The neutral fivebrane is also discussed
in \dufflu.}
The latter two involve finite size
instantons of an unbroken non-Abelian gauge group. For simplicity
we will restrict our analysis to a
generic point in the Narain moduli space where the gauge group is
$U(1)^{28}$ and there will be no finite size gauge instantons. Thus
only the neutral fivebrane can be relevant to our analysis.

According to the
low-energy analysis of \chs\ the neutral fivebrane has $(1,0)$
world-brane supersymmetry with a single hypermultiplet of zero
modes. The hypermultiplet consists of 4 real scalar moduli
associated to translations in the four dimensions transverse to
the brane and a six-dimensional Weyl fermion. The fermion zero
modes arise from the action of the 8 components of $N=1$ supersymmetry
in $d=10$ which are broken by the fivebrane background.

The 8 fermion zero modes are precisely what we require to
get an  instanton-induced 8 fermion interaction
term.  It is important to check that the
collective coordinate integral is well defined.
We do this as follows.
{}From the formulae of \chs\ it is easy to
write down the fermion zeromodes:
\eqn\frzer{
\eqalign{
\delta \lambda  & = 2 e^{-\phi} \Gamma^m \p_m \phi \epsilon \otimes \eta \cr
\delta \psi_\mu & =  - \delta_{\mu m} \p_n \phi \Gamma^{mn} \epsilon \otimes
\eta \cr}
}
where $\mu,m=1,\dots 4$,
$\epsilon \otimes \eta$ is a constant
spinor in the $(2^+,4)$ of $SO(4) \times SO(6)$ and
$\phi$ is given by \chs:
\eqn\fei{
e^{ 2 \phi} = e^{2 \phi_0} + { \alpha' \over  x^2} .
}
Thus the
gauge coupling
diverges ``down the throat''
of the neutral fivebrane. Nevertheless,
the fermion zero modes are normalizable
and localized near the throat (at distances scales $\sim \sqrt \alpha'$).
Moreover, the 8-fermion term inducing
the $R^2$ interactions can be extracted from
equation (2.11) of \bderoo. One finds several
different tensor structures, which can be denoted
schematically as:
\eqn\fermstruct{
\eqalign{
(\bar \psi  \Gamma^{(1)} \psi )^4 \quad \qquad
(\bar \psi  \Gamma^{(1)} \psi )^3 (\bar \psi  \Gamma^{(3)} \psi )
 \quad
&
(\bar \psi  \Gamma^{(1)} \psi )^3 (\bar \psi  \Gamma^{(5)} \psi )
 \quad\qquad \cr
(\bar \psi  \Gamma^{(1)} \psi )^3 (\bar \psi  \Gamma^{(7)} \psi )
\quad \qquad
(\bar \psi  \Gamma^{(1)} \psi )^3
(\bar \psi  \Gamma^{(6)} \lambda )
\quad
 &
(\bar \psi  \Gamma^{(1)} \psi )^3
(\bar \psi  \Gamma^{(4)} \lambda )
\quad\qquad \cr}
}
The notation $\Gamma^{(n)}$ refers to
$\Gamma_{\mu_1\dots \mu_n}$. Indices are contracted
in all possible combinations. All these terms scale in
the same way as $x^2 \rightarrow 0$, and the
density in the collective
coordinate integral behaves like
$$
\int d^4 x {1 \over  x^2}
$$
as $x^2 \rightarrow 0$ so there is no divergence.
The integral also converges well for
$x^2 \rightarrow \infty$.

We can also argue that the weight of the
instanton action  is correct. The wrapped neutral
fivebrane has an action which is $T_5 V_6$ with $T_5$ the fivebrane
tension and $V_6$ the volume of $T^6$. The fivebrane tension
saturates a Bogomolnyi bound given in \andybrane\ and from this bound
the action $T_5 V_6$ is equal to the action of a minimal charge
gauge instanton and is thus equal to
$8 \pi^2 {\rm Re} S$ with our conventions.   We will also
check the action later by comparison to M theory.

\subsec{Wrapped 5brane: microscopic analysis}

So far we have ignored the fact that the fivebrane solutions of
\chs\ have  regions of strong coupling
(``down the throat'')  which can invalidate a naive low-energy analysis
of the zero mode structure and lead to
novel effects \wittensmall. Let us start with the $SO(32)$ heterotic
string in $d=10$. The neutral fivebrane has $(1,0)$ worldbrane supersymmetry
and the zero modes discussed in \chs\ consist of a single neutral
hypermultiplet
whose scalar fields give the location in $R^4$ of the fivebrane. According
to the analysis of \wittensmall\ there are additional non-perturbative
collective coordinates which consist of a $SU(2)$ gauge multiplet with
gauge field ${\cal A}$. There are also
hypermultiplets in the $(2,32)$ of $SU(2) \times SO(32)$.

At a generic
point in the Narain moduli space $SO(32)$ Wilson lines will break the
four-dimensional gauge group to $U(1)^{28}$ and give mass to the
$(2,32)$ hypermultiplets. The fivebrane collective coordinates governing
zero energy deformations of the fivebrane will then consist of
the neutral hypermultiplet plus the values of the flat $SU(2)$
connections  and their fermion partners.

It is convenient  to wrap the fivebrane on
$T^6$ in two steps by regarding $T^6$ as $T^4 \times {\cal T}^2$. From string
duality we know that the Kahler modulus of the ${\cal T}^2$
in this decomposition
is equal to the $S$ modulus of the original
IIA string theory.
We first consider the neutral fivebrane wrapped
on $T^4$. Then as in \wittensmall\ the flat $SU(2)$ connections on $T^4$
are just Wilson lines around the one-cycles $\gamma_i$ of $T^4$
\eqn\wiline{U_i = P \exp \int_{\gamma_i} {\cal A}.}
If $e^{\pm i \theta_i}$ are the eigenvalues of $U_i$ then the moduli space
of flat $SU(2)$ connections has periodic coordinates $\theta_i$ subject
to the Weyl group identification $\theta_i \rightarrow - \theta_i$.
The moduli space of flat connection is thus ${\cal T}^4/Z_2$ where
we use ${\cal T}$ to denote the torus with coordinates $\theta_i$ in
order to distinguish it from the compactification torus $T^4$.
Thus the fivebrane
wrapped on $T^4$ yields a string in ten dimensions with transverse
coordinates propagating on the space $\IR^2 \times \CT^2 \times {\cal
T}^4/Z_2$.
As predicted by string-string duality, this is precisely the structure
of the IIA string compactified on $K3$ as long as it is correct
to view the orbifold ${\cal T}^4/Z_2$ as equivalent to $K3$. We henceforth
refer to ${\cal T}^4/Z_2$ as ${\cal K}3$. This
soliton description  is implicitly in static gauge, but we should be able to
consider the soliton IIA string constructed in this way more abstractly.
We now consider the effects of wrapping the fivebrane on the full
$T^6$. These instantons can be viewed as
worldsheet instantons of the IIA soliton string in
the target space  ${\cal T}^2 \times \CK 3$.
Summing over these instantons will give precisely the same sum
as in the original IIA string theory, but with the replacement
$T \rightarrow \tau_S$.

Thus in this example we have a very direct mapping from second
quantized mirror symmetry not only between
terms in the Lagrangian but also between explicit instanton configurations.
In the original IIA theory we have world-sheet instantons which are
genus one holomorphic curves on $K3 \times T^2$. In heterotic theory these
map to spacetime instanton effects which can be viewed as world-sheet
instantons of the soliton IIA string given by genus one holomorphic
curves on ${\cal K}3 \times {\cal T}^2$. We expect that this point
of view will be useful also in $N=2$ dual pairs. In this case we
can start with world-sheet instantons of the fundamental IIA string
on a Calabi-Yau space which is a $K3$ fibration. On the heterotic
side we have a dual pair consisting of the heterotic string on
$K3 \times T^2$ with a specific
choice of gauge bundle. Indeed, if the
$K3$ surface is elliptically fibered, as in
$F$-compactification \effcomp\ we may
attempt to use the adiabatic argument of
\vwdual\ and  write  the ``fibration'':
\eqn\fibration{
\matrix{
\tilde T^2 \times T^2 & \rightarrow & K3 \times T^2 \cr
  &  &  \downarrow\cr
  &   &  \IP^1 \cr}
}
whose generic fiber is a 4-torus. Once again we can consider
fivebrane
instantons wrapped on $K3 \times T^2$  in a two-step process.
In the first step we wrap the fivebrane on
a generic fiber ${\tilde T}^2 \times T^2$
to obtain a soliton IIA string.
This soliton string propagates on the  ``fibration'':
\eqn\fibrationii{
\matrix{
  \CK 3 & \rightarrow & X_3  \cr
  &  &  \downarrow\cr
  &   &  \IP^1 \cr}
}
giving a Calabi-Yau 3-fold $X_3$,
  where the ${\cal K}3$ fiber
is constructed as before as the
moduli space of flat $SU(2)$ connections
on ${\tilde T}^2 \times T^2$.
Worldsheet instantons where the soliton IIA string wraps the $\IP^1$ will then
give rise to nonperturbative
spacetime instanton effects in the heterotic string.
The adiabatic argument will need to be corrected,
since, for example,   the ${\cal K}3$  fibers degenerate over
$\IP^1$ as do the  ${\tilde T}^2$ fibers  over $\IP^1$.
Nevertheless, it should be  possible again to
map directly the spacetime instantons to the
worldsheet instantons of the dual soliton IIA string on a
${\cal K}3$ fibration
Calabi-Yau space. It is an interesting  problem to determine
the relation between these two Calabi-Yau spaces and to figure out
how the data of the heterotic gauge bundle is encoded in this
description. The answer is provided, in part, by
F-compactification \effcomp.

Even in the $N=4$ context discussed here there are several aspects of this
identification which deserve further
investigation. As in \wittensmall\ we have ignored effects which may
be associated with cancellation between
$SU(2)$ and $SO(32)$ Wilson lines.
However the general picture seems robust
and should continue to hold true whether
or not quantum effects modify the orbifold
${\cal T}^4/Z_2$ to a smooth  $K3$
surface, as discussed in
 \wittensmall\
\ref\divine{N. Seiberg, ``IR Dynamics on
Branes and Space-Time Geometry,'' hep-th/9606017}.

\newsec{Comments on M theory fivebranes}

It is clear that the fivebrane instanton effects described above
must have a description in M theory since the $SO(32)$ heterotic string
is $T$-dual to the $E_8 \times E_8$ theory which can be obtained
from $M$ theory on $S^1/Z_2$ \horwitta.

{}From an analysis of the fermion zero modes it is clear that in M theory
the required fivebrane cannot be the bulk fivebrane with $(2,0)$
worldbrane supersymmetry but must instead be a zero size gauge fivebrane
which lives at the boundary of the Higgs and Coulomb branches
of the M theory fivebrane moduli space. This is the tensionless
string theory when considered in $\IR^{1,5}$ \refs{\edtob, \andyopen,\gh,\swts}
but here compactified in Euclidean signature
on $T^6$.

We can perform one small   check on the M theory
description by computing
the action of the instanton directly in M theory. On general grounds this
must give the same answer as before.

Following the conventions of
\dealwis\ the M theory fivebrane tension is given
in terms of the eleven dimensional Planck constant by
\eqn\fbranten{T_5^M = {\pi^{1/3} \over 2^{1/3} \kappa_{11}^{4/3} } .}
On the other hand it was shown in \horwittb\ that the $E_8$ gauge
coupling $\lambda$ in M theory obeys the relation
\eqn\eightcoupl{\kappa_{11}^4 = { \lambda^6 \over 4 (2 \pi)^5 } .}
Combining this with \fbranten\ gives
\eqn\fnalrel{T_5^M V_6 = {4 \pi^2 V_6 \over \lambda^2 } = {8 \pi^2 \over
(g)^2 } }
where the final factor of two arises from the fact that the
normalization used in \horwittb\ for the gauge kinetic term
differs by a factor of two from the normalization we use in
which the instanton action is $8 \pi^2/g^2$.

\newsec{Conclusions}

We have used string-string duality to compute the $S$ dependent
corrections to $R^2$ couplings in $N=4$ heterotic string theory.
These are given by an infinite set of spacetime instanton
corrections and we have identified the instanton as the neutral fivebrane
of heterotic string theory or equivalently the zero size
fivebrane of the M theoretic description of heterotic string
theory. We have also argued that there is in this example a
direct map from the world-sheet instantons of the IIA
string on $K3 \times T^2$ to spacetime instantons in the heterotic string
consisting
of world-sheet instantons of the IIA soliton string on a dual
$K3 \times T^2$. This direct map from worldsheet instantons to
spacetime instantons viewed as worldsheet instantons of a soliton string
is likely to have application to other dual pairs involving
only $N=2$ or $N=1$ spacetime supersymmetry.

\bigskip

\centerline{\bf Acknowledgements}\nobreak

We would  like to thank T. Banks, O. Ganor, A. Losev,
E. Martinec, S. Shatashvili,
A. Strominger,  and S. Shenker for discussions.
We are extremely grateful to J. Louis for extensive
correspondence on gravitational couplings and
holomorphic anomalies and to V. Kaplunovsky
for an important discussion on this topic.

GM would like to
thank  the Aspen Center for Physics for providing
a stimulating atmosphere during the beginning
of this work. This work was  supported in part
by NSF Grant No.~PHY 91-23780 and
DOE grant DE-FG02-92ER40704.

\listrefs
\bye